\documentclass[pre,aps,twocolumn,superscriptaddress]{revtex4-1}
\usepackage{amsmath,amssymb,graphicx}
\usepackage{epsfig}
\usepackage{textcomp}

\usepackage{subfigure}
\usepackage{graphicx}
\usepackage{url}
\usepackage{float}
\usepackage{cases}
\usepackage[normalem]{ulem}

\usepackage{color}

\begin{document}

\title{
Complete Integrability of the Problem of Full Statistics of Nonstationary Mass Transfer in the Simple Inclusion Process}
\author{Eldad Bettelheim}
\email{eldad.bettelheim@mail.huji.ac.il}
\affiliation{Racah Institute of Physics, Hebrew University of
Jerusalem, Jerusalem 91904, Israel}
\author{Baruch Meerson}
\email{meerson@mail.huji.ac.il}
\affiliation{Racah Institute of Physics, Hebrew University of
Jerusalem, Jerusalem 91904, Israel}
\begin{abstract}
The Simple Inclusion Process (SIP) interpolates between two well-known lattice gas models: the independent random walkers and the Kipnis-Marchioro-Presutti model.  Here we study large deviations of nonstationary mass transfer in the SIP at long times in one dimension. We suppose that  $N\gg 1$ particles start from a single lattice site at the origin, and we are interested in the probability $\mathcal{P}(M,N,T)$
of observing $M$ particles, $0\leq M\leq N$,
to the right of the origin at a specified time $T\gg 1$.
At large times, the corresponding full probability distribution has a large-deviation behavior, $-\ln \mathcal{P}(M,N,T) \simeq \sqrt{T} s(M/N,N/\sqrt{T})$. We determine the rate function $s$ exactly by uncovering and utilizing complete integrability, by the inverse scattering method, of the underlying equations of the macroscopic fluctuation theory. We also analyze different asymptotic limits of the rate function $s$.

\end{abstract}
\maketitle

\section{Introduction}
\label{intro}

The full statistics of  mass or energy transfer (otherwise called integrated current) in macroscopic systems of interacting particles out of equilibrium has been a focus of attention in statistical mechanics in the past two decades. A minimal set of models for detailed studies of this quantity is stochastic lattice gases \cite{Spohn,Liggett,KL,Krapivskybook}. A great deal of progress has
been achieved  in determining the full statistics of mass transfer for nonequilibrium steady states, see \textit{e.g.} Refs. \cite{Derrida2007,BlytheEvans,AppertRolland,Lecomte}.
Nonstationary regimes proved to be much harder for analysis, and exact results for the full statistics of the integrated current (or of the closely related tagged particle position) here are quite
limited \cite{DG2009a,DG2009b,IMS2017,IMS2021,BSM2022a,Grabsch2022,Mallick2022,BSM2022b,KLD2023}.
Exact results were obtained for the Symmetric Exclusion Process (SEP) (see, \textit{e.g.} Ref.~\cite{Spohn}) and the  Kipnis-Marchioro-Presutti (KMP) model \cite{KMP}. Here we focus on a bosonic counterpart of the SEP: the Symmetric Inclusion Process (SIP), and study its full non-stationary mass-transfer statistics. The SIP, first introduced in Ref. \cite{Giardina2010}, describes particles which perform independent symmetric random walks. In addition, each particle ``invites" any other particle, located at a nearest-neighbor position,  to join it on its site, and the invitation is always accepted. The resulting ``inclusion jumps" create an attractive inter-particle interaction. For comparison, the inter-particle interaction in the SEP is repelling, because a particle is not allowed to jump to a site already occupied by another particle.

We consider a system of $N$ particles which start at $t=0$ from a single lattice site at the origin $x=0$.  The particles will spread along the lattice because of the random walk and inclusion jumps.  As the process is stochastic, the number of particles $M$ to the right of the origin will fluctuate in time around the expected value $N/2$. We focus on the fluctuating \emph{excess} number of particles $K=N/2-M$ to the right of the origin at some observation time $t=T$. Our main goal is to determine  $\mathcal{P}(K,N,T)$, the probability of $K$  at time $T$, in the limit of $N\gg 1$ and $T\gg 1$. In the continuum limit, the corresponding probability distribution has a compact support,  $|K|\leq N/2$. It is intuitively clear that,
at large $N$ and $T$, the probability distribution $\mathcal{P}(K,N,T)$ must have a large-deviation form, as it is unlikely to observe nonzero $K$.

A similar non-stationary large-deviation problem for the KMP model has been recently solved  exactly \cite{BSM2022a,BSM2022b}, see also Ref. \cite{KLD2023}. (The KMP model involves immobile agents which occupy a whole lattice and can carry continuous amounts of energy. At each random move the combined energy of a randomly chosen pair of nearest neighbors is randomly redistributed among them.) The solution in Refs. \cite{BSM2022a,BSM2022b} combined two formalisms: the macroscopic fluctuation theory (MFT) and the inverse scattering method (ISM). The MFT (see Ref. \cite{JonaLasinioreview} for a review) is a weak-noise large-deviation formalism rooted in fluctuational hydrodynamics  \cite{Spohn,KL,LL}. The ISM (see \textit{e.g.} Ref. \cite{Novikov}) is a method of solving classical integrable nonlinear equations. It relies on an auxiliary scattering problem which in effect make  these equations linear.

Here we apply the MFT and ISM to the SIP. The MFT formulation reveals the large-deviation scaling of the probability distribution:
\begin{equation}\label{ldscaling}
-\ln \mathcal{P}(K,N,T) \simeq \sqrt{T} s\left(\frac{K}{N},\frac{N}{\sqrt{T}}\right),\quad \sqrt{T}\to \infty,
\end{equation}
and our goal is to calculate the rate function $s$.  We find that the Hopf-Cole transformation brings the MFT equations for the mass transfer statistics to the form of the derivative nonlinear Schroedinger equation (DNLSE) in imaginary space and time. The latter equation can be exactly solved by the Zakharov-Shabat ISM adapted for the DNLS \cite{KaupNewell} and extended  to boundary-value problems \cite{BSM2022a,BSM2022b}. We obtain the rate function $s$ exactly and also analyze its different asymptotic limits.

In Sec. \ref{MFT} we present the governing equations of the fluctuational hydrodynamics and the MFT. Then we perform the Hopf-Cole transformation which brings the MFT equations to
the DNLSE. In Sec. \ref{solution} we employ the ISM to determine the rate function $s$. We briefly summarize and discuss our findings in Sec. \ref{discussion}. In the Appendix we derive four different asymptotics of the rate function. These include very small and very large deviations of $K$, as well as the limits of very small and very large effective densities $n=N/\sqrt{T}$, where our results agree with those for the independent random walkers (RWs) and the KMP model, respectively.

\section{Fluctuational Hydrodynamics and MFT}
\label{MFT}

Exploiting the large parameters $N$ and $\sqrt{T}$, we can describe the system by
fluctuational hydrodynamics. For a general one-component stochastic lattice gas fluctuational hydrodynamics involves a single Langevin equation: a stochastic partial differential equation for the coarse-grained
particle density $\rho(x,t)$:
\begin{equation}\label{Langevin}
\partial_t \rho = \partial_{x} \left[D(\rho)\partial_x \rho +\sqrt{\sigma(\rho)}\,\xi(x,t)\right],
\end{equation}
where $\xi(x,t)$ is white Gaussian noise
\begin{equation}\label{whitenoise}
\langle \xi(x_1,t_1) \xi(x_2,t_2) \rangle = \delta(x_1-x_2) \delta(t_1-t_2)\,.
\end{equation}
Particular lattice gases are specified by the two transport coefficients $D(\rho)$ and $\sigma(\rho)$.
With a convenient choice of units one obtains for the SIP $D = 1$ and $\sigma(\rho) = 2 \rho (1+\rho)$.
For comparison, $D=1$ and $\sigma(\rho) = 2\rho(1-\rho)$ for the SEP, $D=1$ and $\sigma(\rho) = 2\rho^2$ for the Kipnis-Marchioro-Presutti (KMP) model, and $D=1$ and $\sigma(\rho) = 2\rho$ for the independent random walkers (RWs) \cite{Spohn}.
As one can see, the SIP interpolates between the RWs and the KMP models. That is, the SIP behaves as the independent RWs at $\rho \to 0$, whereas at very large densities it becomes similar to the KMP model.

The initial condition for the density is
\begin{equation}\label{incond1}
\rho(x,t=0) = N \delta(x)\,.
\end{equation}
At the observation time $t=T$ we have
\begin{equation}\label{constraint1}
\frac{1}{N}\int_0^{\infty} \rho(x,t=T)\,dx -\frac{1}{2} = \kappa\,,
\end{equation}
where $\kappa \equiv K/N$ is the relative excess of transferred mass with respect to its expected zero value. Obviously, $|\kappa|\leq 1/2$.

Let us rescale the variables, $x/\sqrt{T} \to x$ and $t/T\to t$. For the SIP, the rescaled Langevin equation (\ref{Langevin}) is
\begin{equation}\label{Langevin1}
\partial_t \rho = \partial_{x} \left[\partial_x \rho +T^{-1/4}\sqrt{\rho(1+\rho)}\,\xi(x,t)\right]\,.
\end{equation}
The MFT equations are obtained via the saddle-point approximation to the exact path integral for Eq. ~(\ref{Langevin1}) subject to the condition~(\ref{constraint1}). The saddle-point approximation relies on the small parameter $T^{-1/4}$ in the noise term. As this parameter goes to zero, the probability density $\mathcal{P}(K,N,T)$ becomes dominated by the optimal (that is, most likely) gas density history $\rho(x,t)$ and the ``conjugate momentum"  density history $p(x,t)$ which is closely related to the optimal history of the noise $\xi(x,t)$. A derivation of the MFT equations for the integrated current can be found in many papers, see \textit{e.g.} Refs. \cite{DG2009b,BSM2022a}.  The dynamics of the rescaled $\rho(x,t)$ and $p(x,t)$ are described by  Hamilton's equations
\begin{eqnarray}
  \partial_t \rho &=& \partial_{xx} \rho-2\partial_x \left[\rho(1+\rho) \partial_x p\right], \label{qpa} \\
  \partial_t p &=& - \partial_{xx} p+(2 \rho+1)(\partial_x p)^2\,, \label{qpb}
\end{eqnarray}
The rescaled initial condition (\ref{incond1}) is
\begin{equation}\label{incond1a}
\rho(x,t=0) = n \delta(x)\,.
\end{equation}
The parameter $n=N/\sqrt{T}$ describes the characteristic particle density at the observation time $t=T$. As we will see, it plays an important role in the solution.

The rescaled ``final" condition (\ref{constraint1}) becomes
\begin{equation}\label{constraint1a}
\frac{1}{n}\int_0^{\infty} \rho(x,t=1)\,dx -\frac{1}{2} = \kappa\,.
\end{equation}
It can be accounted for via a Lagrange multiplier which leads to the following ``final" condition on $p$:
\begin{equation}\label{fincond2}
p(x,t=1) = \lambda \theta(x)\,,
\end{equation}
where $\theta(x)$ is the Heaviside's function. The Lagrange multiplier $\lambda$ is ultimately determined by the constraint (\ref{constraint1a}).

The large-deviation scaling of the probability distribution $\mathcal{P}(K,N,T)$, announced  in Eq.~(\ref{ldscaling}), follows directly
from the rescaling transformation $x/\sqrt{T} \to x$ and $t/T\to t$.  The rate function
$s(\kappa,n)$ in Eq.~(\ref{ldscaling}) is given by the rescaled mechanical action of the Hamilton's system (\ref{qpa}) and (\ref{qpb}):
\begin{equation}\label{action}
    s(k,n)=\int_0^1 dt \int_{-\infty}^{\infty} dx\, \rho(1+\rho) (\partial_x p)^2\,.
\end{equation}
It is much simpler, however, to calculate $s(k,n)$ by using the ``shortcut relation" $ds/d\kappa =n \lambda$, see \textit{e.g.} Ref. \cite{shortcut}. This calculation only requires the knowledge of the density profile at $t=1$ as a function of $\lambda$, or the inverse relation.

Nonstationary MFT equations, similar to Eqs. (\ref{qpa}) and (\ref{qpb}), have appeared in many large-deviation problems for diffusive lattice gases. Such coupled nonlinear partial differential equations are usually unamenable to exact solution, and one should resort to numerics and asymptotic limits. Until now only a small number of exceptions have been found \cite{BSM2022a,Mallick2022,BSM2022b,KLD2023}. The list remains quite short even if we add to it recent exact results obtained in a different physical context of full short-time statistics of height as described by the KPZ equation \cite{KLD2021,KLD2022}.

As we show here, the present problem presents us with one more fortunate exception. In order to see it, let us perform the Hopf-Cole canonical transformation from $\rho(x,t)$ and $p(x,t)$ to the new variables
\begin{equation}\label{HC}
u=\rho e^{-p}\,,\quad P=e^p\,\,.
\end{equation}
In the new variables $u(x,t)$ and $P(x,t)$ the Hamilton's equations (\ref{qpa}) and (\ref{qpb}) are
\begin{eqnarray}
  \partial_t u &=& \partial_{xx} u-2\partial_x (u^2\partial_x P), \label{QP1a} \\
  \partial_t P &=& - \partial_{xx} P-2u(\partial_x P)^2. \label{QP1b}
\end{eqnarray}
Remarkably, Eqs.~(\ref{QP1a}) and (\ref{QP1b}) in the new variables formally coincide with the MFT equations for the KMP model in the original variables \cite{DG2009b,BSM2022a,BSM2022b}.

The  boundary conditions in time, Eqs.~(\ref{incond1a}) and (\ref{fincond2})  become
\begin{equation}\label{incond2}
u(x,t=0)=  \nu \,\delta(x)\,,\quad \text{where}\quad \nu=\frac{n}{P(0,0)}\,,
\end{equation}
and
\begin{equation}\label{fincond3}
P(x,t=1)=e^{\lambda \theta(x)}\,.
\end{equation}

Now let us introduce $v(x,t) = -\partial_x P(x,t)$. Equations (\ref{QP1a}) and (\ref{QP1b}) become
\begin{eqnarray}
  \partial_t u  &=& \partial_{x} \left(\partial_x u +2u^2 v\right), \label{Qeq} \\
  \partial_t v &=&  \partial_{x}\left(-\partial_x v+2u v^2\right).\label{veq}
\end{eqnarray}
They should be solved with the initial condition (\ref{incond2}) and the final condition
\begin{equation}\label{fincond4}
v(x,t=1)=-\Lambda \,\delta(x)\,,
\end{equation}
where we have introduced $\Lambda=e^{\lambda}-1$. Using Eq.~(\ref{fincond3}), we can rewrite the integral constraint~(\ref{constraint1a}) in the new variables:
\begin{equation}\label{constraint2}
\frac{1}{n}\int_0^{\infty} u(x,t=1)\,dx =
\frac{\frac{1}{2}+\kappa}{1+\Lambda}\,.
\end{equation}
The MFT problem obeys a  nontrivial symmetry relation in the new variables:
\begin{align}
\nu v(x,t)=-\Lambda u(-x,1-t)\,.
\label{symmetry}
\end{align}

Remarkably, Eqs.~(\ref{Qeq}) and (\ref{veq}) are equivalent to the derivative nonlinear Schroedinger equation (DNLSE) in imaginary space and time \cite{DNLSE}. Formally, they coincide with the MFT equations for the KMP model. This fact was already exploited in Refs. \cite{BSM2022a,BSM2022b} (see also Ref. \cite{KLD2023}) for exactly solving  by the ISM a similar MFT problem for the KMP model.  A complication of the present case is in that, because of the Hopf-Cole transformation, the initial condition~(\ref{incond1a}) becomes ``spoiled" by the presence of an \textit{a priori} unknown quantity $P(x=0,t=0)$.
As we show here, this complication can be overcome in a relatively straightforward manner leading us to the exact rate function $s(\kappa,n)$. At this stage we only note that $P(x=0,t=0)$ can be expressed as an integral of $v(x,t=0)$ over $x$:
\begin{equation}\label{P00}
P(0,0)=1-\int_{-\infty}^0 v(x,0)\,dx\,,
\end{equation}
or alternatively
\begin{equation}\label{P002}
P(0,0)=1+\Lambda+\int_{0}^{\infty} v(x,0)\,dx\,.
\end{equation}

\vspace{0.5cm}
\section{Solution of the problem by the ISM}
\label{solution}

\subsection{Adapting the Kaup-Newell procedure of the ISM}
\label{ISM}

The problem posed by Eqs.~(\ref{Qeq}) and~(\ref{veq}) with boundary conditions in time (\ref{incond2}) and (\ref{fincond4}) can be solved using the ISM. The derivation proceeds along the lines of Ref. \cite{BSM2022a} with some adjustments that we will discuss shortly.  One defines an auxiliary scattering problem for a wave function $\psi(x,t;k)$,
where $k$ is an auxiliary parameter, called `the spectral parameter'. The function $\psi$ satisfies the  evolution equations $\partial_{x}\psi =U{\psi},$  $\partial_{t}\psi =V{\psi},$ where $U$ and $V$ are given by the following expressions:
 \begin{widetext}
\begin{align}
&U=\begin{pmatrix}
-i k/2 & -i   v\sqrt{  i  k} \\
-i  u\sqrt{ i  k} &   i k/2 \\
\end{pmatrix},\\
&V=\begin{pmatrix} k^2/2- i  kuv & -i (\sqrt{ i  k})^3v+ i  \sqrt{ i  k } \,\partial_x v-i  \sqrt{ i  k} 2 v^2u \\
-i (\sqrt{ i  k})^3u+i  \sqrt{ i  k }\, \partial_x u-i  \sqrt{ i  k} 2 u^2v &   -k^2/2+ i kuv, \\
\end{pmatrix}  .
\end{align}
\end{widetext}
As one can check, the compatibility condition $\partial_{t} \partial_{x}{\psi} = \partial_{x}\partial_{t}{\psi}$, which corresponds to the equation
\begin{equation}
\label{eq:compatibilityUV}
\partial_{t}U-\partial_{x}V+\left[U,V\right] = 0
\end{equation}
applied to all $k$, is equivalent to our MFT equations~(\ref{Qeq}) and (\ref{veq}).
The scattering problem $\psi$ consists in finding $\psi(x,t)$ for given $t$ and for $x\to\infty$, given its behavior at $x\to-\infty$. In particular, given that, as $x\to-\infty$, the function $\psi$ is given by
\begin{align}
\psi(x\to -\infty,t)\to  \begin{pmatrix} \alpha e^{-\frac{i kx  }{2}}\\
\beta e^{\frac{i kx  }{2}{}} \\
\end{pmatrix},
\end{align}
one searches for a $2\times 2$ matrix $T(k,t)$ such that  at $x\to\infty$
the function $\psi$ is given by
\begin{align}
\psi(x\to \infty,t)\to\begin{pmatrix}e^{-\frac{i kx  }{2}} & 0 \\
0 & e^{\frac{i kx  }{2}} \\
\end{pmatrix}T(k,t)\begin{pmatrix} \alpha\\
\beta \\
\end{pmatrix}.
\end{align}
The elements of the matrix $T(k,t)$ can be written as follows:
\begin{align}
T(k,t)=\begin{pmatrix}a(k) & \tilde{b}(k)e^{k^2t} \\
{b}(k)e^{-k^2t} & \tilde{a}(k) \\
\end{pmatrix}\,.
\end{align}
As one can see, the diagonal elements are time-independent, while the off-diagonal ones have a very simple $e^{\pm k^2t}$ time dependence. The reason for this simple form is the simplified form of the matrix $V$ at $x\to\pm\infty$ because all the fields vanish there.

One can find the expressions for  $T(k,t)$ at $t=0$ and $t=1$ in terms of the fields $v(x,0)$ and $u(x,1)$, respectively, by making use of the boundary conditions in time for Eqs.~(\ref{Qeq}) and~(\ref{veq}).
Indeed, due to these boundary conditions, the equation $\partial_{x}\psi =U{\psi}$ becomes rather simple, and the result can be given by the Fourier transform of the fields $v(x,0)$ and $u(x,1)$. In fact, one can compute $\tilde b(k)$ in two different ways -- by using the data either at $t=0$ or at $t=1$ -- and obtain the following  equation:
\begin{align}
\!-i \sqrt{ i  k}\left[Q(k)- i  k\nu Q_{-}(k) Q_{+}(k)\right]\!=\!\tilde b(k)\!=\!i \lambda e^{-k^2}\sqrt{ik},\label{beq}
\end{align}
where
\begin{eqnarray}
  \!Q_+(k) &\!=\!& \int_0^\infty e^{i k x}v(x,0)dx,\;\;Q_+(k)\!=\!\int_{-\infty}^0 e^{i k x}v(x,0)dx, \nonumber \\
 &&\quad\quad\quad Q(k)=Q_+(k)+Q_-(k).
\end{eqnarray}
Equation~(\ref{beq}) can be recast in the form
\begin{align}
\left[1-i k\nu Q_+(k)\right]\left[1-i k\nu Q_-(k)\right]=1+i  \lambda \nu ke^{-k^2}.
\end{align}
Note that the factors $1-i k\nu Q_\pm(k)$ are analytic in the upper or lower half planes, respectively. Therefore, the logarithm of the left hand side is a  Wiener-Hopf decomposition of the logarithm of the right hand side. Since the Wiener-Hopf decomposition is achieved by a Cauchy integral, we can write down a solution for $Q_\pm(k)$ in terms of an exponent of that integral:
\begin{eqnarray}
\!\!&&\!\!1-i k \nu Q_\pm(k) \nonumber \\
\!\!&&\!\!= \left(1\pm\nu   v_{\pm}\right)\exp\left[\pm\int_{-\infty}^{\infty}\!\frac{\ln\left(1+i\lambda k'e^{-k'^{2}}\right)}{k'-k\mp i0^{+}}\frac{dk'}{2\pi i }\right],
\end{eqnarray}
where $v_\pm$ are constants to be determined from the demand that $Q_\pm(k)$ be regular at the origin. These constants arise from the ambiguity of the Wiener-Hopf decomposition with respect to an additive constant (which is exponentiated here to form a multiplicative constant).

The demand that $Q_\pm(k)$ is regular at the origin, ensuring a well-behaved $v(x,0)$ at infinity, yields:
\begin{align}
1\pm\nu   v_\pm=\exp\left[\mp\int_{-\infty}^{\infty}\frac{\ln\left(1+i\lambda k'e^{-k'^{2}}\right)}{k'\mp i0^{+}}\frac{dk'}{2\pi i }\right].
\end{align}
Since $Q_\pm$ is the Fourier transform of $v(x,0)\theta(\pm x)$, one can obtain $v(x,0)$\ by an inverse Fourier transform.
The result is:
\begin{widetext}
\begin{equation}
v(x,0)=\frac{1}{\nu}  \,\int_{-\infty}^{\infty}\frac{e^{-i k x}}{i k}\Bigg\{1-\exp \left[\pm\int_{-\infty}^{\infty}\left(   \frac{\ln \left(1+i \Lambda \nu k' e^{-k'^2}\right)}{k'-k\mp i 0^+}-\frac{\ln \left(1+i \Lambda \nu k' e^{-k'^2}\right)}{k'}\right)\frac{dk'}{2\pi i }\right]\Bigg\}dk,
\label{IST1}
\end{equation}
\end{widetext}
where $\pm$ in this equation is equal to the sign of $x$.

\subsection{Calculating the rate function $s(\kappa,n)$}
\label{algebra}

Using Eq.~(\ref{IST1}), we obtain $v_\pm=v(0^\pm,0)$ as follows:
\begin{align}
\mp\nu  v_\pm=1-\exp\left[\mp\int_{-\infty}^{\infty}\frac{\ln \left(1+i \Lambda \nu k e^{-k^2}\right)}{k}\frac{dk}{2\pi i }\right]
\end{align}
Now we use Eq.~(\ref{P00}) for $P(0,0)$:
\begin{equation}
  P(0,0) \!-\!1= \frac{1}{4\pi\nu}\!\int_{-\infty}^{\infty}\!\frac{\ln \left(1+\Lambda^2 \nu^2 k^2e^{-2k^2}\right)}{ k^2}\,dk+\frac{\Lambda}{2}.
 \label{P00eq}
\end{equation}
Since $\nu = n/P(0,0)$, Eq.~(\ref{P00eq}) is actually a transcendental equation which relates $P(0,0)$ to $\Lambda$ at given $n$. This equation can be rewritten as
\begin{align}
\int_{-\infty}^{\infty}\frac{ \ln \left(1+ \frac{\Lambda^2 n^2}{\mu^2} k^2e^{-2k^2}\right)}{k^2}\frac{dk}{4\pi}=\frac{n}{\mu}\left(\mu-1-\frac{\Lambda}{2}\right)\,.
\label{eq1}
\end{align}
where we have denoted $P(0,0) =\mu$ for brevity. Now we turn to  condition~(\ref{constraint2}), for which we need to compute $\int_0^{\infty} u(x,1)\,dx$. By virtue of the symmetry relation (\ref{symmetry}), we have
\begin{equation}\label{massright}
\int_0^{\infty} u(x,1)\,dx = -\frac{\nu}{\Lambda} \int_{-\infty}^0 v(x,0)\,dx = \frac{\nu}{\Lambda}\left(\mu-1\right)\,,
\end{equation}
where the second equality uses Eq.~(\ref{P00}). Then Eq.~(\ref{constraint2}) yields
\begin{equation}\label{eqmu}
\mu=\mu(\kappa,\Lambda)=\frac{2(1+\Lambda)}{2-2\kappa \Lambda+\Lambda}\,.
\end{equation}
Plugging it into Eq.~(\ref{eq1}), we arrive at a single
equation for $\Lambda$ at given $\kappa$ and $n$:
\begin{equation}
\label{eq2a}
\int_{-\infty}^{\infty}\frac{\ln \left(1+ \frac{\Lambda^2 n^2}{\mu^2} k^2 e^{-2k^2}\right)}{k^2}\frac{dk}{4\pi} =
\frac{\Lambda  n \left[2 \kappa  (\Lambda +2)-\Lambda \right]}{4 (\Lambda +1)}\,,
\end{equation}
where $\mu$ is given by Eq.~(\ref{eqmu}).

\begin{figure}[ht]
\includegraphics[width=0.3\textwidth,clip=]{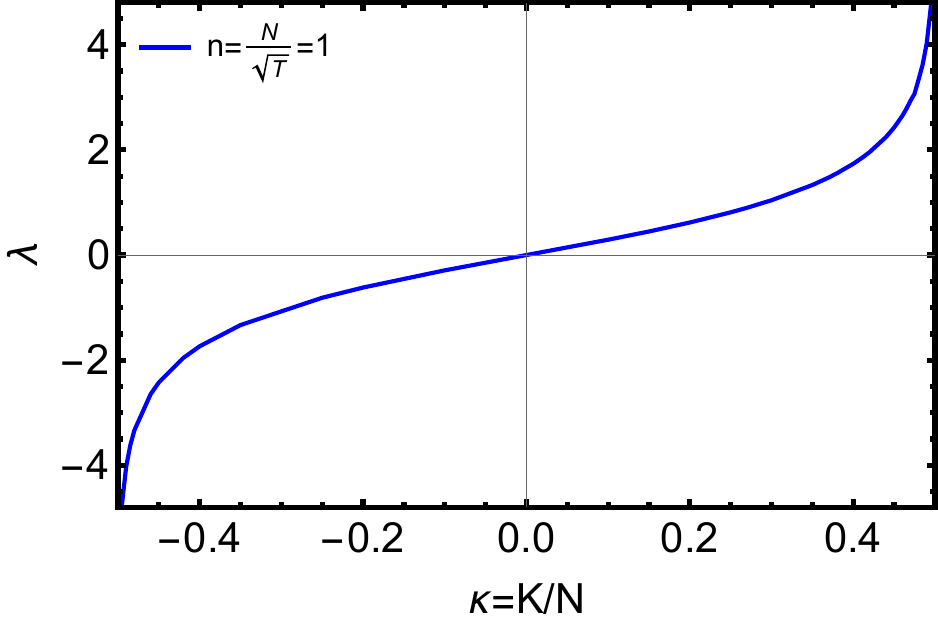}
\caption{The Lagrange multiplier $\lambda=\ln(1+\Lambda)$ vs. the relative excess of transferred mass $\kappa=K/N$, found by numerically solving the transcendental equation (\ref{eq2a}) for $n=N/\sqrt{T}=1$.}
\label{lambdavskappa}
\end{figure}

Equation~(\ref{eq2a}) is invariant under the transformation $\kappa \to -\kappa$ and $\Lambda \to -\Lambda/(1+\Lambda)$, which corresponds to the antisymmetry relation  $\lambda(-\kappa) = -\lambda(\kappa)$. In view of the shortcut relation $ds/d\kappa = n \lambda$, this antisymmetry relation reflects the physically obvious symmetry $s(-\kappa) = s(\kappa)$ of the rate function.

Equation (\ref{eq2a}) can be solved numerically for $\lambda\equiv\ln(1+\Lambda)$ as a function of $\kappa$ at any given effective particle density $n$. An example of such a solution is shown in Fig. \ref{lambdavskappa} for $n=1$. With the function $\lambda(\kappa,n)$ at hand, one can compute the rate function by integrating $\lambda(\kappa,n)$ over $\kappa$: $s(\kappa,n)=n\int_0^{\kappa} \lambda(\kappa',n) d\kappa'$. Because of the reflection symmetry of the problem, $x\to -x$,
it suffices to consider $0\leq \kappa\leq 1/2$.

\begin{figure}[ht]
\includegraphics[width=0.3\textwidth,clip=]{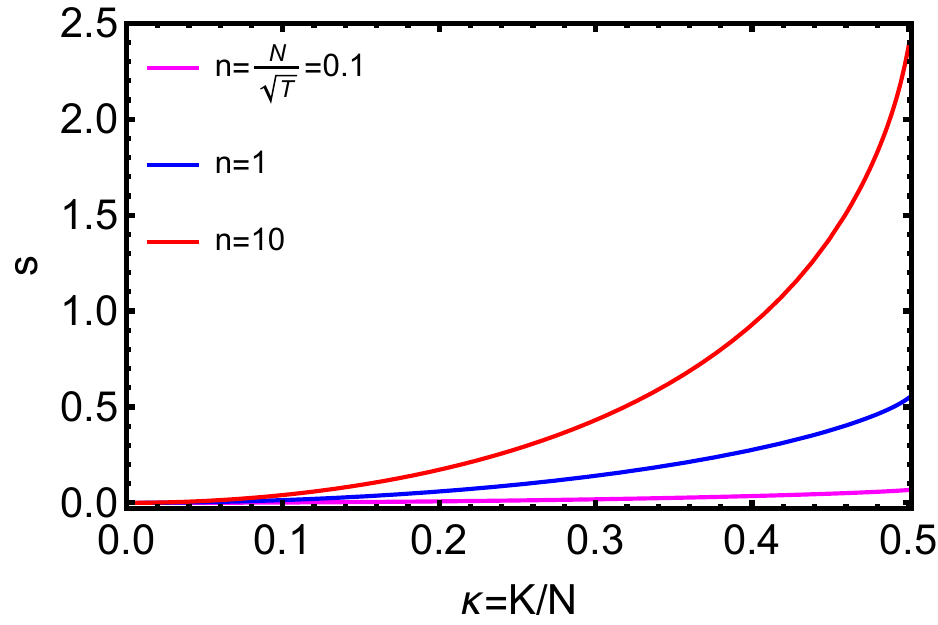}
\caption{The rate function $s(\kappa,n)$ vs. the relative excess of transferred mass $\kappa$ for $n=0.1$, $1$ and $10$.}
\label{threeactions}
\end{figure}

Figure \ref{threeactions}
compares the resulting rate function $s(\kappa,n)$ versus $\kappa$ for three different values of $n$. At fixed $\kappa$ the rate function grows with $n$, as to be expected on physical grounds.

\begin{figure}[ht]
\includegraphics[width=0.3\textwidth,clip=]{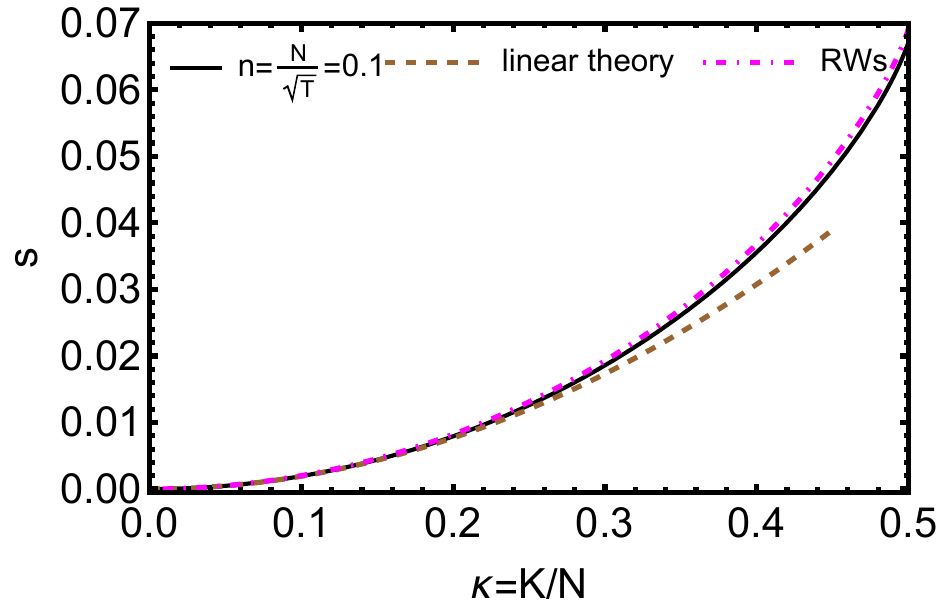}
\caption{The rate function $s(\kappa,n)$ vs. the relative excess of transferred mass $\kappa$ for $n=0.1$. Also shown, for the same $n=0.1$, the $\kappa\ll 1$ asymptotic and the rate function for the RWs.}
\label{n01}
\end{figure}

\begin{figure}[ht]
\includegraphics[width=0.3\textwidth,clip=]{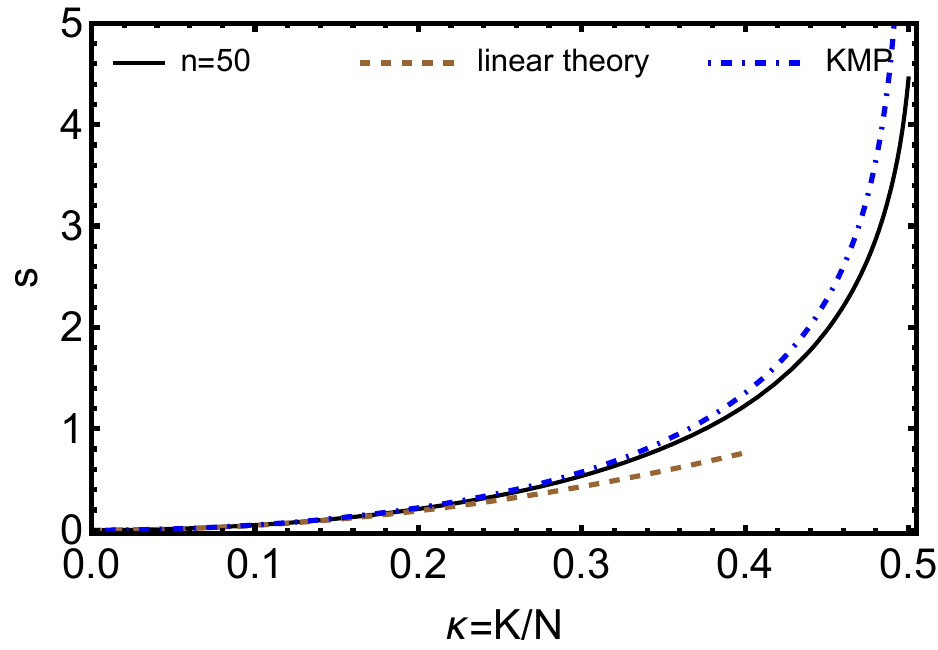}
\caption{The rate function $s(\kappa,n)$ vs. relative excess of transferred mass $\kappa$ for $n=50$. Also shown, for the same $n=50$, the $\kappa\ll 1$ asymptotic  and the rate function for the KMP model \cite{BSM2022a}.}
\label{n50}
\end{figure}

\begin{figure}[ht]
\includegraphics[width=0.3\textwidth,clip=]{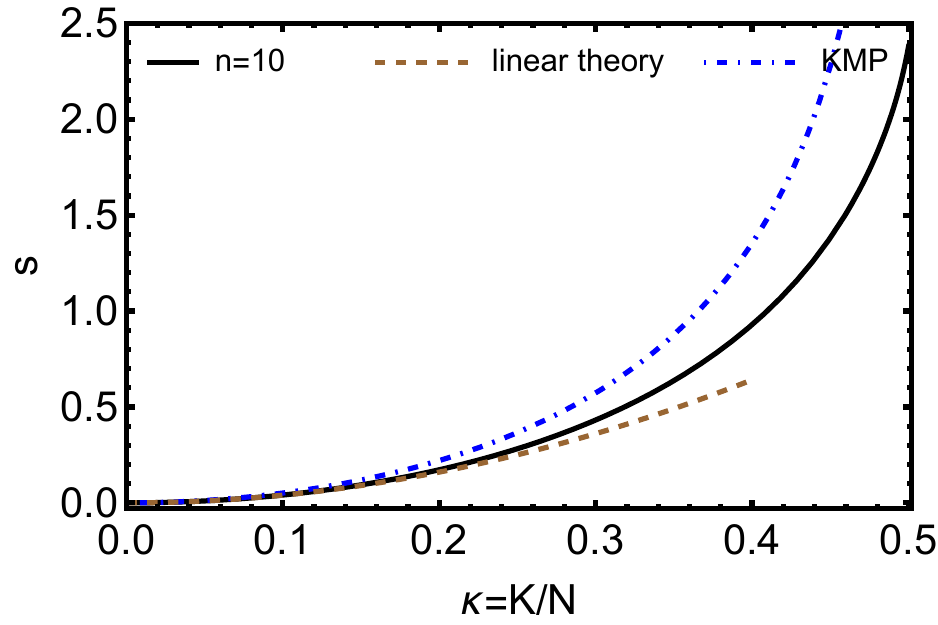}
\caption{The rate function $s(\kappa,n)$ vs. $\kappa$ for $n=10$. Also shown, for the same $n=10$, the $\kappa\ll 1$ asymptotic  and the rate function for the KMP model \cite{BSM2022a}.}
\label{n10}
\end{figure}

It is instructive to consider several asymptotic regimes of the exact rate function $s(n,\kappa)$. One of them is the asymptotic at $\kappa\ll 1$ which is shown, for different values of $n$, in Figs.~\ref{n01}-\ref{n10}. This asymptotic describes typical, small fluctuations of  $\kappa$, and it can be obtained from the linear theory \cite{KrMe}, see section A of the Appendix.

Another interesting asymptotic describes $s(n,\kappa)$ close to the edges of support of the distribution, $\kappa = \pm 1/2$. Here the SIP and the KMP models behave very differently.
For the KMP model the probability distribution of $\kappa$ vanishes at $\kappa = \pm 1/2$, so the corresponding rate function diverges at these points \cite{BSM2022a}. As we elaborate in section B of the Appendix, for  the SIP the rate function remains finite at $\kappa = \pm 1/2$, as is the case for the RWs.

As we mentioned earlier, the SIP interpolates between two lattice gas models: the noninteracting random walkers (RWs)  and the KMP model. Therefore, one can expect that, at $n\ll 1$ and $n\gg 1$ the rate function $s(\kappa,n)$ for the SIP should approach that for the RWs  and for the KMP model, respectively. These properties are indeed observed in Figs. \ref{n01} (for $n=0.1$) and \ref{n50} (for $n=50$), respectively. For comparison, for an intermediate value of $n=10$  the predicted rate function for  the KMP model is still considerably higher than that for  the SIP (Fig. \ref{n10}). In section C of the Appendix we present exact solution of the mass excess problem for the RWs and compare it with the $n\to 0$ limit of the SIP. Then, in section D of Appendix, we show how the KMP limit arises from Eq.~(\ref{eq2a}) at large $n$.

\section{Summary and Discussion}
\label{discussion}

We employed the ISM to determine exactly the full long-time statistics of mass transfer, for a localized initial condition, of the SIP. The resulting large-deviation rate function $s(\kappa,n)$ interpolates in a nontrivial way between two lattice gas models -- the RWs and the KMP model, for each of which the exact mass transfer statistics has been known. This interpolation is controlled by the parameter $n=N/\sqrt{T}$, the effective density of the system at the observation time $T$. In particular, there is a difference in behavior between the SIP, even with an arbitrary large effective density $n=N/\sqrt{T}$, and the KMP model: At $\kappa = \pm 1/2$  the rate function  remains finite for the former, but diverges for the latter. This fundamental difference can be traced down to the different nature of the microscopic models: a continuous energy variable in the KMP model versus discrete particles in the SIP and RWs.

The MFT is as a particular example of the optimal fluctuation method (OFM), a universal and versatile tool for studying large deviations of macroscopic systems.
Revealing and exploiting exact integrability of the OFM equations in different contexts
offers pathway toward making substantial progress in some of them.

\subsection*{Acknowledgments}
We are grateful to Ohad Shpielberg for a useful discussion of the SIP.
Our research is supported by the US-Israel
Binational Science Foundation through Grant No. 2020193 (E.B.) and by the Israel
Science Foundation through Grant 1499/20 (B.M.). B.M. also acknowledges support from the Simons Foundation Targeted Grant 920184 to the Fine Theoretical Physics Institute at the University of Minnesota.


\section*{Appendix. Asymptotics of the rate function $s(\kappa,n)$}
\label{asymptotics}

\appendix

\subsection*{A. $\kappa\ll 1$: linear theory}
\label{lintheory}
\renewcommand{\theequation}{A\arabic{equation}}
\setcounter{equation}{0}
For small $\kappa$ Eqs.~(\ref{qpa}) and (\ref{qpb}) can be linearized with respect to $|\lambda|$ \cite{KrMe}:
\begin{eqnarray}
  \partial_t \rho &=& \partial_{xx} \rho , \label{qlinear} \\
  \partial_t p &=& - \partial_{xx} p . \label{plinear}
\end{eqnarray}
As a result, the rate function (\ref{action}) becomes in the leading order
\begin{equation}\label{actionlinear}
    s_{\text{lin}}=\int_0^1 dt \int_{-\infty}^{\infty} dx\, \rho_0(x,t)\left[1+\rho_0(x,t)\right] \left[v(x,t)\right]^2 ,
\end{equation}
where
\begin{equation}\label{rhov}
 \rho_0(x,t) = \frac{n e^{-\frac{x^2}{4 t}}}{\sqrt{4 \pi t}}\quad\text{and}\quad v(x,t) = -\frac{\lambda e^{-\frac{x^2}{4 (1-t)}}}{\sqrt{4\pi(1-t)}}\,.
\end{equation}
are the solutions of Eqs.~(\ref{qlinear}) and (\ref{plinear}) with the corresponding initial or final condition, respectively. Notice that $\rho_0(x,t)$ describes the deterministic (zero-noise) evolution of the system. Plugging Eqs.~(\ref{rhov}) into Eq.~(\ref{actionlinear}) and evaluating the double integral, we obtain
\begin{equation}\label{actionlinear1}
s_{\text{lin}}(\lambda)=\frac{1}{16} \lambda ^2 n
   \left(\sqrt{\frac{2}{\pi }}
   n+2\right)\,.
\end{equation}
Using the ``shortcut relation" $ds/d\kappa=n\lambda$, we finally obtain
\begin{equation}\label{actionlinear2}
s=\frac{4n  \kappa ^2}{\sqrt{\frac{2}{\pi }} n+2}\,.
\end{equation}
Exactly the same expression (\ref{actionlinear2}) is obtained by expanding the left- and right-hand sides of the exact equations (\ref{eq2a}) and (\ref{eqmu}) at small $\Lambda$ (which in this limit is equal to $\lambda$). To this end one should expand the logarithm in the numerator of the integrand,
\begin{equation}\label{logexp}
\ln \left(1+ \frac{\Lambda^2 n^2}{\mu^2} k^2 e^{-2k^2}\right) \simeq \frac{\Lambda^2 n^2}{\mu^2} k^2 e^{-2k^2}\,,
\end{equation}
and evaluate the resulting integral over $k$.

In each of the limits of $n\to 0$ and $n\to \infty$ Eq.~(\ref{actionlinear2}) coincides with the corresponding $\kappa\ll 1$ asymptotics for the RWs and the KMP model, see sections C and D of this  Appendix, respectively. For the RWs the distribution  of $K$ is independent of time, and the
variance, back in the original variables, is equal to $N/4$, as we also show in section C of this Appendix.

\subsection*{B. $\kappa\to 1/2$}
\label{edge}
\renewcommand{\theequation}{B\arabic{equation}}
\setcounter{equation}{0}
When $\kappa$ approaches $1/2$,  $\Lambda$ goes to $\infty$, and Eq.~(\ref{eq2a}) becomes, in the leading order,
\begin{equation}
\label{eq2aasympt}
\int_{-\infty}^{\infty}\frac{\ln \left[1+ n^2 (1+\Lambda \delta)^2 k^2 e^{-2k^2}\right]}{k^2}\frac{dk}{4\pi} =
\frac{n}{2} (1-\Lambda \delta)\,,
\end{equation}
where $\delta = 1/2-\kappa \ll 1$. As one can see, $\Lambda$ and $\delta$ appear
only through the combination $\Lambda \delta$. Therefore, the solution of Eq.~(\ref{eq2aasympt}) has the form $\Lambda \simeq f(n)/\delta$, with a function $g(n)$ that can be found numerically. Since $\Lambda=e^{\lambda}-1\gg 1$, this is equivalent to
$\lambda\simeq \ln [f(n)/\delta]$. Now it is evident from the relation $s=n\int_{0}^{1/2} \lambda (\kappa',n) \,d\kappa'$ that $s$ remains finite at $\kappa=1/2$, as indeed observed in Figs. \ref{threeactions}-\ref{n10}. At $\kappa$ close to $1/2$ the rate function behaves as
\begin{equation}\label{close05}
s(\kappa,n) \simeq s_0(n) - n \left(\frac{1}{2}-\kappa\right) \ln \frac{f(n)}{\frac{1}{2}-\kappa}\,.
\end{equation}

\subsection*{C. $n\ll 1$: Noninteracting Random Walkers}
\label{RWdelta}
\renewcommand{\theequation}{C\arabic{equation}}
\setcounter{equation}{0}
Here we consider the complete mass transfer statistics in the system of $N \gg 1$ independent random walkers (RWs), simultaneously released at the origin at $t=0$. At long times this model becomes identical to that of independent Brownian particles. Let us start with a microscopic solution, which is very simple. At $t>0$ the expected number of particles in the region $x>0$ is equal to $N/2$.
We are interested in the probability $\mathcal{P}$ that,  at time $t=T$, there are exactly $M$ particles at $x>0$, where $0\leq M\leq N$. The probability that a \emph{single} particle is found at $x>0$ is $1/2$; it is independent of time $T$. Since the RWs are independent,  the probability $\mathcal{P}$ we are after is given by the binomial distribution
\begin{equation}\label{binomial}
P(M,N) ={N\choose M}\, 2^{-N}\,.
\end{equation}

\begin{figure}[ht]
\includegraphics[width=0.25\textwidth,clip=]{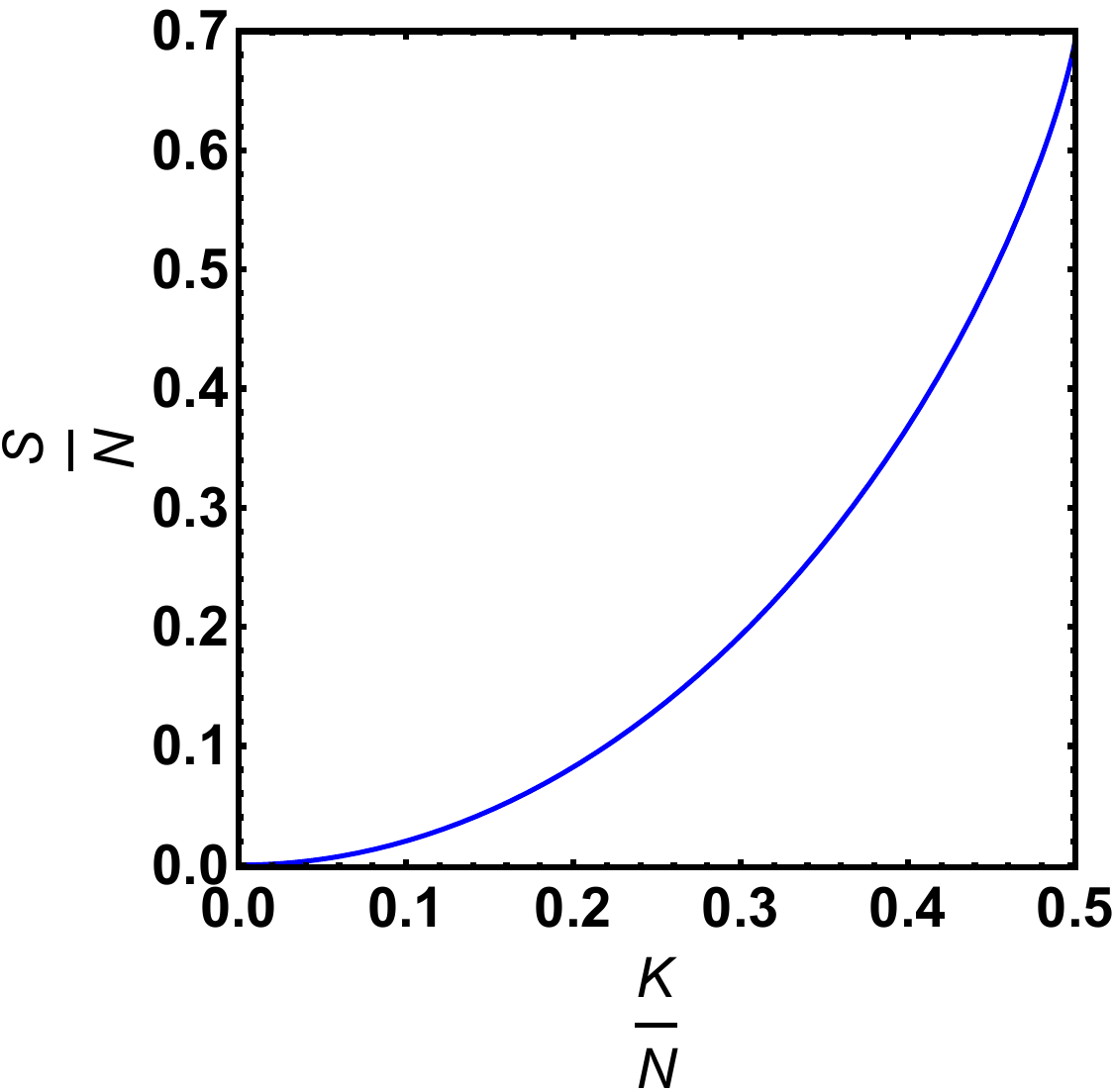}
\caption{The rate function $s=S/N$ vs. $\kappa=K/N$ for the RWs.}
\label{SRW}
\end{figure}

Now we assume that $N,M\gg 1$ and solve the same problem by using the MFT. In this way  we can also determine the optimal path of the system conditioned on a specified $K$.  The MFT equations for the RWs are, in the original variables:
 \begin{eqnarray}
  \partial_t \rho &=& \partial_x (\partial_x \rho -2\rho \partial_x p), \label{RW1} \\
  \partial_t p &=& -\partial^2_x p -(\partial_{x}p)^2. \label{RW2}
\end{eqnarray}
These equations coincide with the $\rho \to 0$ limit of Eqs.~(\ref{qpa}) and (\ref{qpb}) for the SIP.  The initial condition is $\rho(x,t=0) = N \delta(x)$, and the condition at $t=T$ is $p(x,T) = \lambda \theta(x)$, where $\lambda$ is the Lagrange multiplier, and $\theta(x)$ is the theta-function.
The action, $-\ln \mathcal{P} \simeq S$, takes the form
\begin{equation}
  S  = \int_0^T dt \int_{-\infty}^{\infty} dx \,\rho (\partial_x p)^2. \label{actionRW1}
\end{equation}
This MFT problem is exactly solvable via the Hopf-Cole transformation (\ref{HC}). In the new variables one obtains two decoupled linear equations
\begin{eqnarray}
  \partial_t u &=& \partial_{xx} u\,, \label{equ} \\
  \partial_t P &=& - \partial_{xx} P\,, \label{eqP}
\end{eqnarray}
with the boundary conditions in time
\begin{equation}\label{BCRW}
u(x,0) P(x,0) = N \delta(x)\,,\quad P(x,T)=e^{\lambda \theta(x)}\,.
\end{equation}
This problem can be solved first for $P(x,t)$ and then for $u(x,t)$. In particular, we obtain, already in the original variables, the optimal density history
\begin{equation}\label{qrw}
 \rho(x,t) = \frac{N e^{-\frac{x^2}{4 t}} \left[\left(e^{\lambda
   }-1\right) \text{erf}\left(\frac{x}{2
   \sqrt{T-t}}\right)+e^{\lambda }+1\right]}{
   \sqrt{4\pi t} \left(e^{\lambda }+1\right)}\,.
\end{equation}
The Lagrange multiplier $\lambda$ can be found from the condition
\begin{equation}\label{excessRW}
\frac{1}{N}\int_0^{\infty} \rho(x,T)\,dx -\frac{1}{2} = \frac{K}{N} \equiv \kappa\,,
\end{equation}
where again $K=M-N/2$ is the excess number of particles at $x>0$. Equation~(\ref{excessRW}) yields
\begin{equation}\label{lambdaRW}
\lambda = 2\, \text{arctanh} \left(2\kappa\right),
\end{equation}
Using Eq.~(\ref{lambdaRW}) and the relation $dS/d\kappa= N\lambda$, we obtain the rate function $s(\kappa)\equiv S(\kappa,N)/N$:
\begin{equation} \label{srw}
s(\kappa) = \frac{1+2\kappa}{2} \ln \left(1+2\kappa\right)
  +\frac{1-2\kappa}{2} \ln \left(1-2\kappa\right).
\end{equation}
As to be expected, this expression is independent of the measurement time $T$. The ensuing probability  $\mathcal{P}(K,N)\sim \exp[-N s(\kappa)]$  coincides with the $N\gg 1$ asymptotic of the binomial distribution (\ref{binomial}).  Figure \ref{SRW} shows a plot of
$S/N$ vs. $\kappa$.

The quadratic asymptotic of $S$ at $\kappa\ll 1$ describes typical, small fluctuations of $K$, with $\text{Var}_K =N/4$.  The largest possible deviations of $K$ correspond to the maximum values of $S$, $S_{\max}(K=\pm N/2)=N\ln 2$, which are achieved at the edges of the distribution support and agree with the exact microscopic probabilities
$\mathcal{P}(M=0)=\mathcal{P}(M=N) = 2^{-N}$. This probability is manifestly finite.

The optimal density profiles $\rho(x,t)$ at different rescaled times $t/T$, as described by Eqs.~(\ref{qrw}) and~(\ref{lambdaRW}), are shown in Fig. \ref{q(x,t)RW}. Noticeable is a growing with time left-right asymmetry which culminates at $t=T$ as a density discontinuity at $x=0$.

\begin{figure}[ht]
\vspace{0.5cm}
\includegraphics[width=0.35\textwidth,clip=]{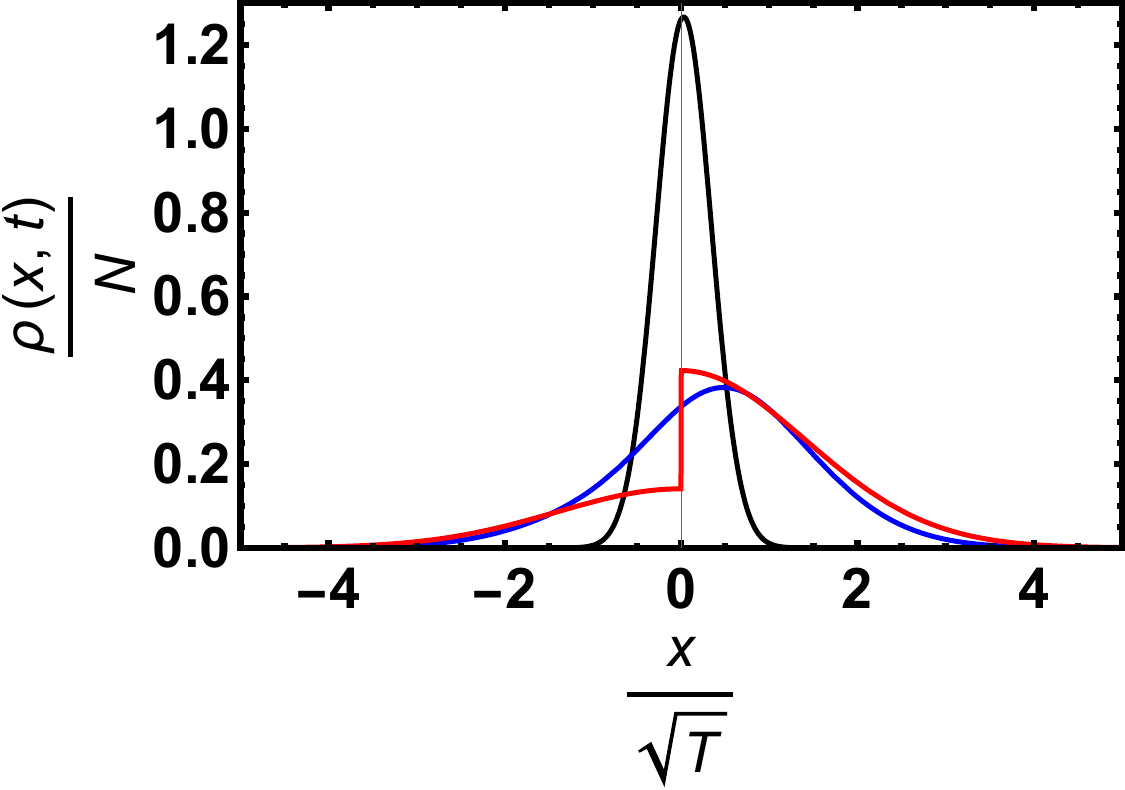}
\caption{The spatial profiles of the optimal density history $\rho(x,t)$  for the RWs for $\kappa=K/N =1/4$ at times $t/T = 0.05$ (black), $0.7$ (blue) and $1$ (red).}
\label{q(x,t)RW}
\end{figure}

Now we show how Eq.~(\ref{lambdaRW}) for the RWs arises in the limit of $n\to 0$ of Eq.~(\ref{eq2a}) for the SIP. At small $n$ the numerator of the integrand in Eq.~(\ref{eq2a}) can again be expanded at small argument, as in Eq.~(\ref{logexp}). Evaluating the resulting integral, we arrive, after cancellations, at the following equation:
\begin{equation}\label{smalln}
\frac{\Lambda n (\Lambda-2 \kappa  \Lambda
    +2)^2}{4 \sqrt{2 \pi }
   (\Lambda +1)} = 2 \kappa  (\Lambda
   +2)-\Lambda \,.
\end{equation}
In the limit of $n\to 0$ the right hand side must vanish, and we obtain
\begin{equation}\label{smallnLambda}
\Lambda = \frac{4\kappa}{1-2\kappa}\,.
\end{equation}
The resulting $\lambda\equiv \ln(1+\Lambda)$ yields Eq.~(\ref{lambdaRW}).

Finally, at $\kappa$ close to $1/2$ the rate function (\ref{srw}) agrees with Eq.~(\ref{close05}) with $s_0= n\ln 2$ and $f(n)=1/2$, independent of $n$.

\subsection*{D. $n \gg 1$: the KMP limit}
 \label{largen}
 \renewcommand{\theequation}{D\arabic{equation}}
\setcounter{equation}{0}
The rate function for the KMP model was calculated  in Ref. \cite{BSM2022a}. Here we
show how it can be recovered from our Eq.~(\ref{eq2a}) for the SIP in the limit of $n\to \infty$.
When $n\to \infty$, and once $\kappa$ is not too close to $1/2$, $\Lambda$ goes to zero in such a way that $\Lambda n=O(1)$, and Eq.~(\ref{eq2a}) yields
\begin{equation}
\label{eq2alargen}
\kappa = \frac{1}{n \Lambda}\int_{-\infty}^{\infty}\frac{\ln \left(1+ \Lambda^2 n^2 k^2 e^{-2k^2}\right)}{k^2}\frac{dk}{4\pi}\,.
\end{equation}
This expression exactly coincides with  Eq. (27) of Ref. \cite{BSM2022a}, once we identify $n\Lambda$ with the Lagrange multiplier $\lambda$ of Ref. \cite{BSM2022a} to account for the slightly different rescalings in Ref.~\cite{BSM2022a} and here.

This KMP-like behavior breaks down in a narrow boundary layer near $\kappa=1/2$ leading to a finite rate function at $\kappa=1/2$, see section B of this Appendix. This boundary layer shrinks to zero in the limit of $n\to \infty$.

\end{document}